\documentclass[letterpaper,english,reprint,aps]{revtex4-1}
\usepackage[T1]{fontenc}
\usepackage[latin9]{inputenc}
\usepackage{textcomp}
\usepackage{amsmath}
\usepackage{amssymb}
\usepackage{graphicx}
\usepackage{esint}

\makeatletter

\pdfpageheight\paperheight
\pdfpagewidth\paperwidth

\newcommand{\lyxmathsym}[1]{\ifmmode\begingroup\def\b@ld{bold}
  \text{\ifx\math@version\b@ld\bfseries\fi#1}\endgroup\else#1\fi}

\makeatother

\usepackage{babel}
\begin{document}
\title{}
\title{Energy balance and energy correction in dynamics of classical spin
systems}
\author{Dmitry A. Garanin}
\affiliation{Department of Physics, Herbert H. Lehman College and Graduate School,
The City University of New York, 250 Bedford Park Boulevard West,
Bronx, New York 10468-1589, USA}
\begin{abstract}
Energy-correction method is proposed as an addition to mainstream
integrators for equations of motion of systems of classical spins.
This solves the problem of non-conservation of energy in long computations
and makes mainstream integrators competitive with symplectic integrators
for spin systems that for different-site interactions conserve the
energy explicitly. The proposed method is promising for spin systems
with single-site interactions for which symplectic integrators do
not conserve energy and thus have no edge against mainstream integrators.
From the energy balance in the spin system with a phenomenological
damping and Langevin fields, a formula for the dynamical spin temperature
in the presence of single-site anisotropy is obtained.
\end{abstract}
\maketitle

\section{Introduction}

As computing capabilities grow, models of classical spins on a lattice
receive unfading attention. They allow description of both magnetic
structures at low temperatures and thermal disordering effects, including
phase transitions. The latter is an advantage compared to the more
traditional approach, micromagnetics, that struggles to incorporate
the temperature. The fastest method to compute the thermodynamics
of magnetic systems is, of course, Monte Carlo. However, more versatile
is the dynamical approach to classical magnetic systems using the
equation of motion \citep{lanlif35} for lattice spins, in which the
temperature can be introduced either via the phenomenological Landau-Lifshitz
damping \citep{lanlif35} and stochastic Langevin fields \citep{bro63pr}
simulating the heat bath or microscopically via the coupling to the
elastic system of the solid.

The stochastic equations of motion for classical spins are usually
solved numerically by the Heun method with a small integration step
$\delta t$ \citep{pallaz98prb} (for a review, see Ref. \citep{evansetal14jpc}).
For this method, the step error is $\delta t^{3}$ and thus the accumulated
error is $\delta t^{2}$. However, in the important case of a weak
coupling to the bath, one can replace the continuous Langevin noise
by the pulse noise \citep{gar17pre} and, between the regular noise
pulses, use more accurate and efficient integrators such as the classical
fourth-order Runge-Kutta (RK4) method (step error $\delta t^{5}$)
or even Butcher's RK5 method having a step error $\delta t^{6}$ (for
a general introduction to ordinary differential equations, see Ref.
\citep{haiwannor93book}; the RK5 code can be found, e.g., in the
Appendix of Ref. \citep{gar17pre}). This allows one to solve the
Landau-Lifshitz-Langevin equation with the same computing speed as
the usual Landau-Lifshitz equation and in particular to efficiently
solve the problem of non-uniform thermal activation of a magnetic
particle considered as a system of many spins \citep{gar18prb_pn,gar18prb_sa}.
The idea of splitting the deterministic and stochastic parts of the
spin motion was proposed earlier \citep{madud11prb} using the Suzuki-Trotter
(ST) decomposition of the evolution operators.

The latter is a part of a major development in computational physics:
implementation of symplectic integrators that have some important
advantages in comparison with classical ordinary differential equations
(ODE) solvers. The main advantage of symplectic methods is explicit
energy conservation for conservative systems. For classical spin systems,
the algorithm consists in sequential rotating spins around effective
fields acting on them. This explicitly conserves the spin length.
If the effective field depends on the other spins, this rotation also
conserves the energy of the system. The energy conservation is very
important. Long computations on conservative systems using non-energy-conserving
solvers cause energy drift that accumulates to significant values
if the integration step is not very small. This can be interpreted
as a positive or negative fictitious damping in the system. Sometimes
instabilities develop in computations, which results in the system
warming up and becoming disordered. This cannot happen if the numerical
method conserves energy.

There are different types of Suzuki-Trotter decomposition of evolution
operators for spin systems \citep{krebunlan98,lanbunevekretsa00,omemryfol00cmp,omemryfol01prl,tsakrelan04bjp,stesch06}.
The simplest second-order Suzuki-Trotter decomposition (ST2) is easy
to program and fast in the execution. Its accuracy is not great, with
a step error $\delta t^{3}$, but the energy conservation makes the
method viable. Accurate treatment of the energy also improves the
accuracy of other physical quantities. This is probably why currently
in most cases the second-order decomposition is used (see, e.g., \citep{mawoodud08prb,beathibar12prb,basvergarkac17jpcm,strungaruetal21prb}).
The fourth-order decomposition (ST4) (step error $\delta t^{5}$)
is computation-intensive and cumbersome to program. Also worth mentioning
is the implicit spherical mid-point rule \citep{frahualei97jcp,maipenros14pre}.

A drawback of symplectic integrators for spin systems is that they
are hardly suitable for systems with single-site interactions, such
as a crystal field. The effective field produced on the spin by the
single-site anisotropy depends on the spin itself and changes as the
spin is precessing around it. Considering this effective field to
be constant and equal to its value for the starting orientation of
the spin leads to nonconservation of energy. The second-order Suzuki-Trotter
decomposition loses one order of accuracy, so the step error becomes
$\delta t^{2}$ and the accumulated error becomes $\delta t$. If
the single-site anisotropy is much smaller than the exchange, this
could be tolerated at short times, but without the exact energy conservation
the approach loses its edge and cannot be called symplectic. The problem
of a nonconstant effective field was solved by iterations \citep{krebunlan98,lanbunevekretsa00},
but this makes the method cumbersome and causes slowdown. This difficulty
had been overcome in a rather unexpected way: Researchers could not
sacrifice the popular numerical method and instead abandoned models
with single-site anisotropy. For pure spin models, an anisotropic
exchange is used instead of the latter. In the models unifying spin
and lattice dynamics, spin-lattice interaction is introduced via the
dependence of the exchange coupling on the distance between the neighboring
atoms, modified by lattice deformations, and/or via the pseudodipolar
coupling, in which the distances and directions are also modified
by phonons (see, e.g., \citep{beathibar12prb}).

The purpose of this work is to rehabilitate the traditional methods
of solving equations of motion for classical spins that have no problems
with single-site interactions. The nonconservation of the spin length,
accumulating at large times, can be easily corrected by normalization
of all spins from time to time. Correcting the energy is less trivial
and it is discussed in detail. The idea is the following. If the expected
energy of the system is known (e.g., in isolated conservative systems
it remains is the same, and in non-isolated systems it increases by
the amount of the absorbed energy and decreases by the amount of the
dissipated energy), one can change the state of the system by a small
amount to compensate for the mismatch between the target (expected)
energy and the actual energy subject to drift as the result of accumulating
numerical errors or slowly developing instability. For the systems
of particles having kinetic energy, the energy correction is quite
simple: It is sufficient to multiply all momenta by a number found
from the condition that the new total energy equals the target energy.
For spin systems a suitable transformation of the state is less trivial
and it is explained in the paper.

The paper is organized as follows. In Sec. \ref{sec:The-model} the
classical spin model with single-site anisotropy interacting with
the environment via the phenomenological damping and stochastic Langevin
fields is introduced. The rate of change of the system's energy due
to all factors is worked out. At equilibrium this renders the formula
for the dynamical spin temperature. The method of energy correction
based on the balance of the energy flow is explained and constructed
in Sec. \ref{sec:The-energy-correction}, the main part of the paper.
The proposed method is tested on a two-spin toy model having an analytical
solution in the limit of small uniaxial anisotropy in Sec. \ref{sec:Toy model}.
Here the long-time dynamics is computed with the help of different
uncorrected and corrected numerical integrators, including RK4, RK5,
as well as ST2, for a comparison. The efficiency of the proposed method
is demonstrated. Concluding remarks are given in the Discussion.

\section{The model and the energy balance}

\label{sec:The-model}

Consider a classical spin system on the lattice described by the Hamiltonian
\begin{equation}
\mathcal{H}=-\frac{1}{2}\sum_{ij}J_{ij}\mathbf{s}_{i}\cdot\mathbf{s}_{j}-\frac{D}{2}\sum_{i}\left(\mathbf{n}_{i}\cdot\mathbf{s}_{i}\right)^{2}-\mathbf{H}(t)\cdot\sum_{i}\mathbf{s}_{i},\label{Ham}
\end{equation}
where $J_{ij}$ is the exchange coupling, $D$ is the uniaxial anisotropy
that can be coherent or random, depending on the directions of the
local anisotropy axes $\mathbf{n}_{i}$, and $\mathbf{H}(t)$ is the
time-dependent magnetic field in energy units. The dynamics of this
system is described by the Landau-Lifshitz-Langevin equation that
phenomenologically accounts for the interaction of spins with a heat
bath:
\begin{equation}
\hbar\dot{\mathbf{s}}_{i}=\mathbf{s}_{i}\times\left(\mathbf{H}_{\mathrm{eff},i}+\boldsymbol{\zeta}_{i}\right)-\alpha\mathbf{s}_{i}\times\left(\mathbf{s}_{i}\times\mathbf{H}_{\mathrm{eff},i}\right).\label{LLL}
\end{equation}
Here the effective field is given by
\begin{equation}
\mathbf{H}_{\mathrm{eff},i}=-\frac{\partial\mathcal{H}}{\partial\mathbf{s}_{i}}=\sum_{i}J_{ij}\mathbf{s}_{j}+D\left(\mathbf{n}_{i}\cdot\mathbf{s}_{i}\right)\mathbf{n}_{i}+\mathbf{H}(t),\label{Heff}
\end{equation}
$\alpha$ is the dimensionless damping constant \citep{lanlif35},
and $\boldsymbol{\zeta}_{i}$ are the Langevin white-noise fields
satisfying
\begin{equation}
\left\langle \zeta_{i\alpha}(t)\zeta_{j\alpha}(t')\right\rangle =2\alpha\hbar T\delta_{ij}\delta_{\alpha\beta}\delta(t-t').\label{Langevin-correlator}
\end{equation}

The time derivative of the system's energy is given by
\begin{equation}
\mathcal{\dot{H}}=-\dot{\mathbf{H}}(t)\cdot\sum_{i}\mathbf{s}_{i}-\sum_{i}\mathbf{H}_{\mathrm{eff},i}\cdot\mathbf{\dot{s}}_{i}.
\end{equation}
Substituting the equation of motion, one obtains
\begin{eqnarray}
\mathcal{\dot{H}} & = & -\dot{\mathbf{H}}(t)\cdot\sum_{i}\mathbf{s}_{i}-\frac{1}{\hbar}\sum_{i}\mathbf{H}_{\mathrm{eff},i}\cdot\left(\mathbf{s}_{i}\times\boldsymbol{\zeta}_{i}\right)\nonumber \\
 &  & \qquad-\frac{\alpha}{\hbar}\sum_{i}\left(\mathbf{s}_{i}\times\mathbf{H}_{\mathrm{eff},i}\right)^{2}.\label{E_balance_0}
\end{eqnarray}
Here the first term is the power input into the spin system by the
time-dependent magnetic field, the second term is the power input
in the system by the heat bath, and the last term is the dissipated
power. In a large system, the second term has to be averaged over
the realizations of the Langevin fields $\boldsymbol{\zeta}_{i}$.
Calculation in the Appendix results in the energy balance equation
\begin{eqnarray}
\mathcal{\dot{H}} & = & -\dot{\mathbf{H}}(t)\cdot\sum_{i}\mathbf{s}_{i}+\frac{\alpha T}{\hbar}\left\{ 2\sum_{i,j}J_{ij}\mathbf{s}_{i}\cdot{\bf s}_{j}\right.\nonumber \\
 &  & +\left.D\left[3\left(\mathbf{n}_{i}\cdot\mathbf{s}_{i}\right)^{2}-1\right]\right\} -\frac{\alpha}{\hbar}\sum_{i}\left(\mathbf{s}_{i}\times\mathbf{H}_{\mathrm{eff},i}\right)^{2}.\label{E_balance}
\end{eqnarray}
The first and last terms on the right-hand side of this equation are
the absorbed power of the applied field and the dissipated power,
\begin{eqnarray}
P_{\mathrm{abs}} & = & -\dot{\mathbf{H}}(t)\cdot\sum_{i}\mathbf{s}_{i}\nonumber \\
P_{\mathrm{diss}} & = & \frac{\alpha}{\hbar}\sum_{i}\left(\mathbf{s}_{i}\times\mathbf{H}_{\mathrm{eff},i}\right)^{2}.\label{P_abs_diss}
\end{eqnarray}
At equilibrium $\dot{\mathbf{H}}(t)=0$ and $\mathcal{\dot{H}}=0$,
so the energy input from the heat bath via the Langevin fields is
equal to the energy dissipated to the heat bath. This implies that
$T=T_{S}$, where $T_{S}$ is the dynamical spin temperature defined
by
\begin{equation}
T_{S}\equiv\frac{\sum_{i}\left(\mathbf{s}_{i}\times\mathbf{H}_{\mathrm{eff},i}\right)^{2}}{2\sum_{i,j}J_{ij}\mathbf{s}_{i}\cdot{\bf s}_{j}+D_{R}\sum_{i}\left[3\left(\mathbf{n}_{i}\cdot\mathbf{s}_{i}\right)^{2}-1\right]}.\label{TS}
\end{equation}
If all spins are aligned with their effective fields, $\mathbf{s}_{i}\times\mathbf{H}_{\mathrm{eff},i}=0$
and thus $T_{S}=0$. If spins are totally disordered, then for a large
system both terms in the denominator average to zero, and $T_{S}=\infty$.
Equation (\ref{TS}) without the single-site anisotropy was obtained
in Ref. \citep{madudsemwoo10pre}, also by the Langevin formalism.
The validity of this formula is more general. For instance, one can
create a spin state by Monte Carlo at the temperature $T$ and check
$T=T_{S}$. In fact, the formula for the dynamical spin temperature
was obtained earlier for the microcanonical ensemble \citep{nursch00pre}
using the ideas developed for hamiltonian systems \citep{rugh97prl,bannur98pre}.
Equation (\ref{TS}) follows from Eq. (15) of Ref. \citep{nursch00pre}
as a particular case.

\section{The energy correction}

\label{sec:The-energy-correction}

Integrating Eq. (\ref{E_balance}), one obtains the time dependence
of the system's energy due to different processes. The integrals of
the three terms on the right-hand side are robust in the numerical
solution. The work done on the system is counted and does not change
with time. On the contrary, the energy on the left-hand side is not
robust and drifts because of the accumulation of numerical errors.
It is especially clear for the isolated conservative system when the
rhs is trivially zero but the lhs is slowly drifting because of numerical
errors if spins are moving and the integrator does not conserve the
energy explicitly. However, if high-accuracy ODE solvers are used,
the energy drift is very small and can be compensated for by the energy-correction
procedure repeated from time to time. This procedure changes the system's
energy by the small amount
\begin{equation}
\delta E=E_{\mathrm{target}}-E,\label{deltaE}
\end{equation}
where $E_{\mathrm{target}}$ is the precice target value of the energy
obtained by integrating the rhs of Eq. (\ref{E_balance}) and $E$
is the imprecise value of the energy subject to drift and determined
from the instantaneous spin state. The proposed change of the spin
state is
\begin{equation}
\delta\mathbf{s}_{i}=\xi\mathbf{s}_{i}\times\left(\mathbf{s}_{i}\times\mathbf{H}_{\mathrm{eff},i}\right),\label{delta_s_i}
\end{equation}
where the factor $\xi$ is chosen so that the energy changes by $\delta E$.
To first order, the change of system's energy is given by
\begin{equation}
\delta E=-\sum_{i}\mathbf{H}_{\mathrm{eff},i}\cdot\delta\mathbf{s}_{i}=\xi\sum_{i}\left(\mathbf{s}_{i}\times\mathbf{H}_{\mathrm{eff},i}\right)^{2},
\end{equation}
where from
\begin{equation}
\xi=\frac{\delta E}{\sum_{i}\left(\mathbf{s}_{i}\times\mathbf{H}_{\mathrm{eff},i}\right)^{2}}.
\end{equation}
The new spins $\mathbf{s}_{i}+\delta\mathbf{s}_{i}$ should be normalized.
This energy-correction method works as a compensative damping or antidamping.
It is efficient if the fictitious damping due to numerical errors
is not too high that is satisfied in high-accuracy computations.

One can ask what the accuracy of the energy-correcting transformation
is and whether it changes the order of the ODE solver. Indeed, reinstating
the energy value, one possibly can sacrifice the accuracy of other
physical quantities. To clarify this, for the step error of the ODE
solver one can write $\delta s_{i}^{(p)}\sim\delta t^{p+1}$, where
$p$ is the accuracy order of the method. The error accumulated over
$n$ integration steps that require the time $t=n\delta t$ is given
by $\delta s_{i,n}^{(p)}\sim n\delta s_{i}^{(p)}\sim t\delta t^{(p)}$.
The energy-correcting transformation of the first order given above
compensates for the first-order term in the energy due to the accumulated
errors in spin vectors. One has $\delta E\sim\delta s_{i,n}^{(p)}$
and then for the energy-correcting spin changes one obtains $\delta s_{i}\sim\xi\sim\delta E\sim\delta s_{i,n}^{(p)}$.
This means that the deformations of trajectories due to the energy-correcting
transformation are of the same order as the accumulated errors, that
is, the order of the ODE solver is not affected.

The restoration of the energy is incomplete as there are also quadratic
terms in the energy expansion: $\delta^{2}E\sim\left[\delta s_{i,n}^{(p)}\right]^{2}$.
These residual terms are very small, especially for high-order integrators
and a not too long interval between the energy-restoring procedures,
$n$. One can do another energy-correcting transformation to eliminate
this term too. Then the residual term would be $\left[\delta s_{i,n}^{(p)}\right]^{4}$.
This iteration procedure converges very quickly. However, in practical
cases, one iteration is sufficient.

The interval between energy corrections depends on the required accuracy
of the energy. If only one iteration is done, the remaining error
in the energy is $\left[\delta s_{i,n}^{(p)}\right]^{2}\sim\left[n\delta s_{i}^{(p)}\right]^{2}$.
For high-order integrators, the step error $\delta s_{i}^{(p)}$is
small, and thus $n$ can be large, which reduces the computing load.
In any case, there is no need to perform the energy correction after
each integration step of the basic ODE solver.

\section{Checking the energy-correcting method for a toy model}

\label{sec:Toy model}

\begin{figure}
\begin{centering}
\includegraphics[width=9cm]{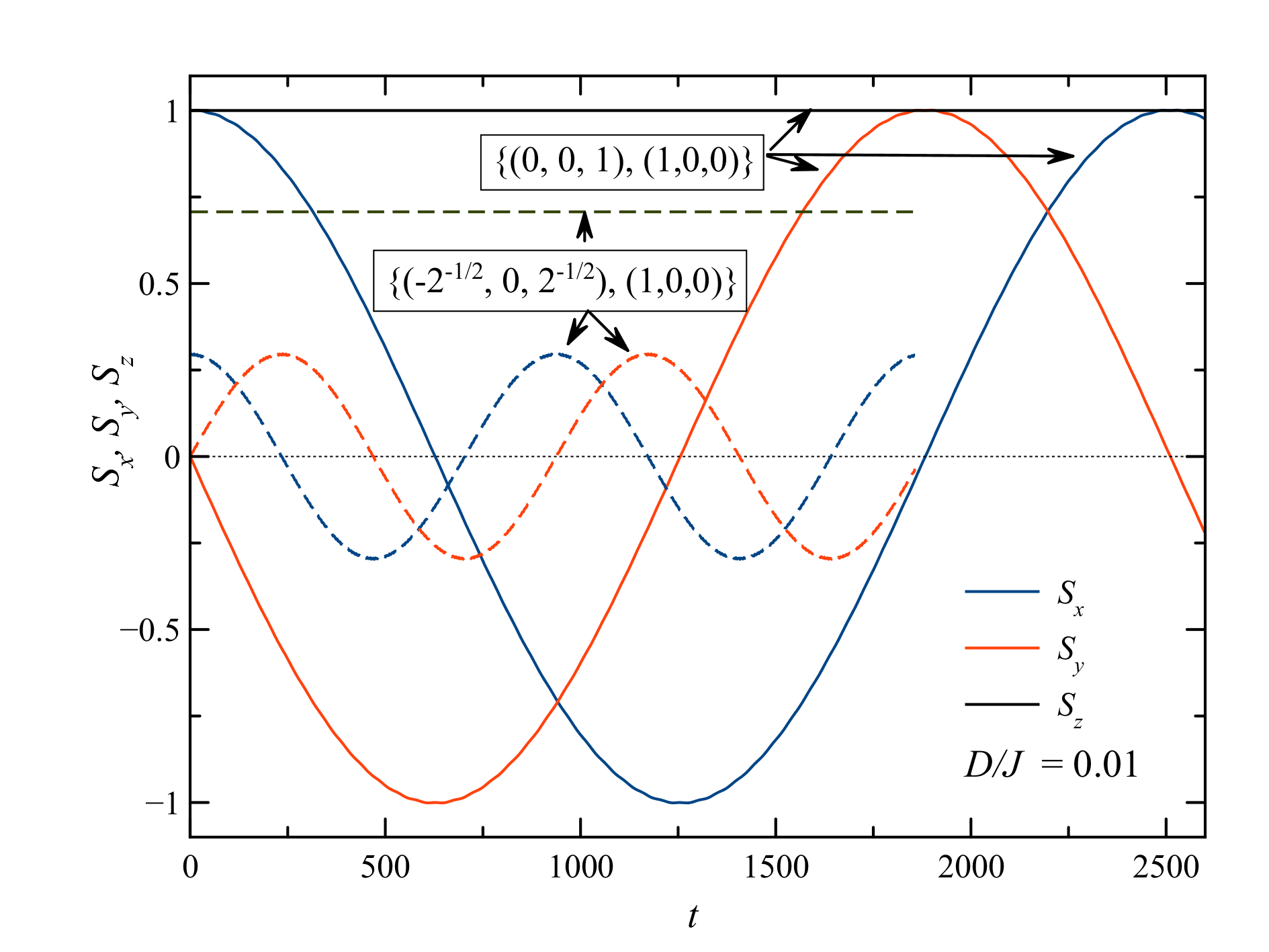}
\par\end{centering}
\caption{Slow precession of the total spin in the model of two coupled spins
with a small uniaxial anisotropy, $D/J=0.01$. The direction of the
precession of the total spin depends on the angle between the two
spins. \label{Fig-Sxyz_vs_t_D=00003D0p01_hStep=00003D0p1_both_cases}}
\end{figure}

To see how the energy-correction method works with mainstream ODE
integrators for classical-spin systems, consider a toy model of two
coupled spins with uniaxial anisotropy
\begin{equation}
\mathcal{H}=-J\mathbf{s}_{1}\cdot\mathbf{s}_{2}-\frac{D}{2}\left(s_{1z}^{2}+s_{2z}^{2}\right).
\end{equation}
The equations of motion for the spins have the form
\begin{eqnarray}
\hbar\mathbf{\dot{s}}_{1} & = & \mathbf{s}_{1}\times H_{\mathrm{eff},1}=\mathbf{s}_{1}\times\left(J\mathbf{s}_{2}+D\mathbf{e}_{z}s_{1z}\right)\nonumber \\
\hbar\mathbf{\dot{s}}_{2} & = & \mathbf{s}_{2}\times H_{\mathrm{eff},2}=\mathbf{s}_{2}\times\left(J\mathbf{s}_{1}+D\mathbf{e}_{z}s_{2z}\right).\label{s1s2_Eqs}
\end{eqnarray}
The state of this system is specified by four angles: $\theta_{1}$,
$\phi_{1}$, $\theta_{2}$, and $\phi_{2}$, There are two integrals
of motion: $\mathcal{H}$ and $S_{z}=s_{1z}+s_{2z}$, thus the equations
of motion can be represented via only two dynamical variables. The
general solutions should be complicated though.

\begin{figure}
\begin{centering}
\includegraphics[width=9cm]{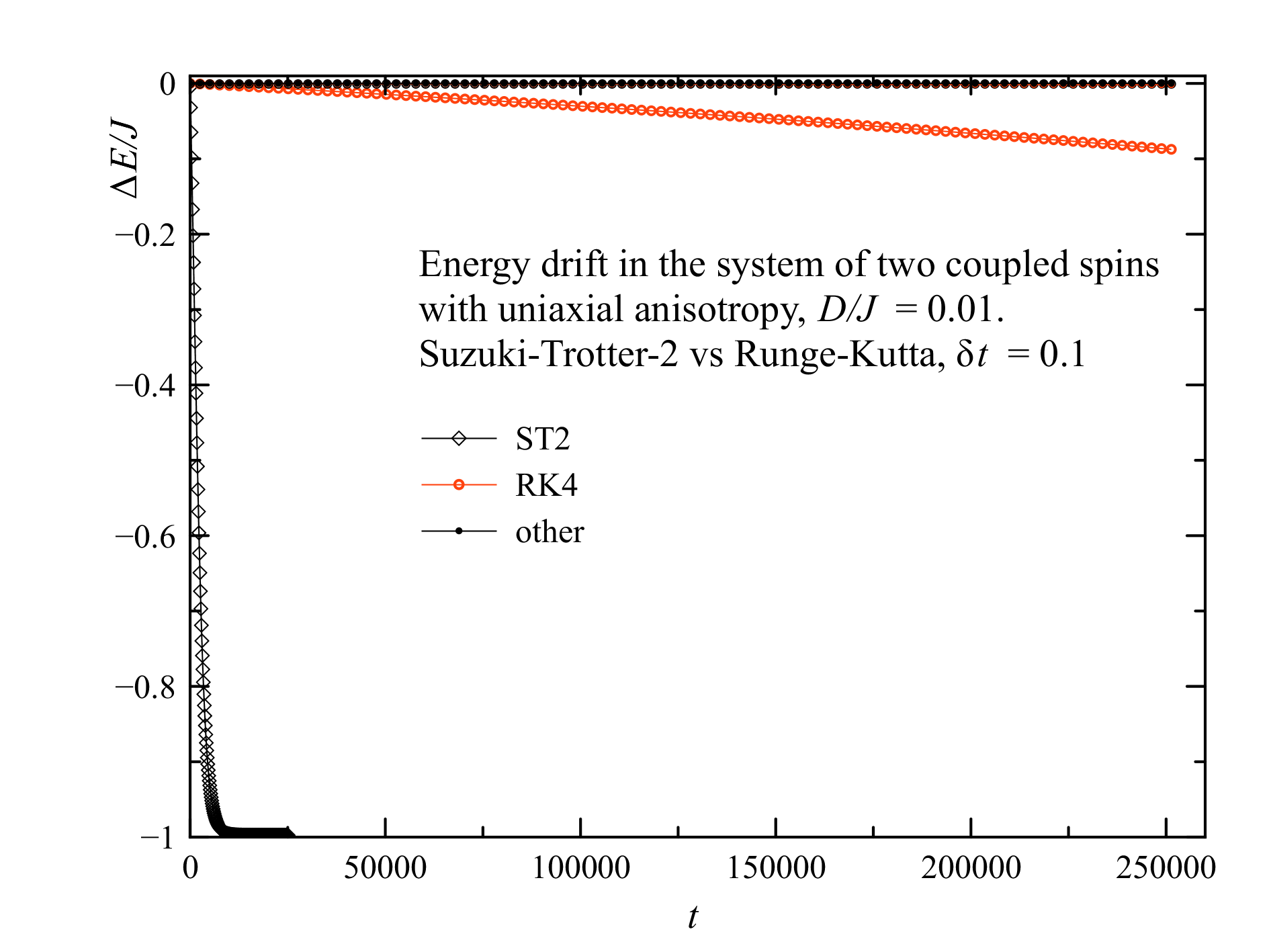}
\par\end{centering}
\caption{Energy drift in the system of two coupled spins with uniaxial anisotropy,
$D/J=0.01$, computed with the ST2 and Runge-Kutta methods with an
integration step $\delta t=0.1$ \label{Fig-DeltaE_vs_t_D=00003D0p01_hStep=00003D0p1_ST2_vs_RK}}

\end{figure}

\begin{figure}
\begin{centering}
\includegraphics[width=9cm]{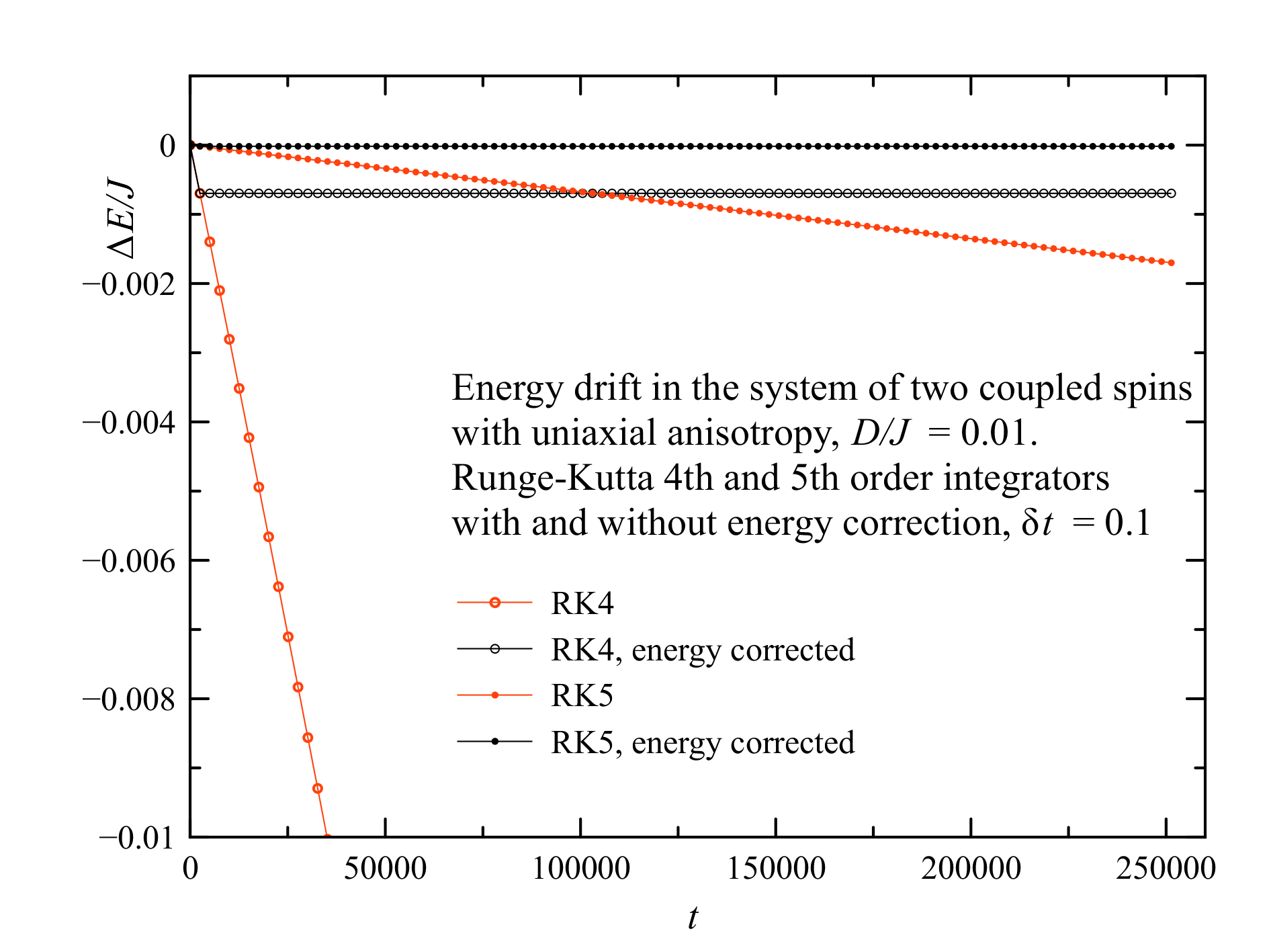}
\par\end{centering}
\caption{Close-up of the energy drift in the system of two coupled spins with
uniaxial anisotropy $D/J=0.01$, computed with Runge-Kutta fourth-
and fifth-order integrators with and without energy correction, with
an integration step $\delta t=0.1$ \label{Fig-DeltaE_vs_t_periods_D=00003D0p01_hStep=00003D0p1_zoom}}
\end{figure}

An approximate analytical solution is possible in the limit $D\ll J$
where there is a fast precession of spins around the total spin and
a slow precession of the total spin around $z$ axis. In terms of
new variables
\begin{equation}
\mathbf{S}=\mathbf{s}_{1}+\mathbf{s}_{2},\qquad\boldsymbol{\sigma}=\mathbf{s}_{1}-\mathbf{s}_{2}
\end{equation}
the equations of motion become
\begin{eqnarray}
\hbar\mathbf{\dot{S}} & = & \frac{1}{2}D\mathbf{S}\times\mathbf{e}_{z}S_{z}+\frac{1}{2}D\boldsymbol{\sigma}\times\mathbf{e}_{z}\sigma_{z}\label{S_Eq}\\
\hbar\boldsymbol{\dot{\sigma}} & = & \frac{1}{2}J\boldsymbol{\sigma}\times\mathbf{S},\label{sig_Eq}
\end{eqnarray}
where in the second equation the small terms with $D$ are discarded.
One can see that the motion of the total spin $\mathbf{S}$ is slow.
In the equation for $\mathbf{S}$, the second term has to be averaged
over the fast precession of $\boldsymbol{\sigma}$ around $\mathbf{S}$.
After some vector algebra one obtains the resulting equation of motion
for the total spin
\begin{equation}
\mathbf{\dot{S}}=\Omega\mathbf{S}\times\mathbf{e}_{z},\qquad\hbar\Omega=\frac{D}{4}S_{z}\frac{1+3\mathbf{s}_{1}\cdot\mathbf{s}_{2}}{1+\mathbf{s}_{1}\cdot\mathbf{s}_{2}}.\label{Omega_res}
\end{equation}
The direction of precession of $\mathbf{S}$ depends not only on $S_{z}$,
but also on the angle $\theta_{12}$ between the two spins. For $\mathbf{s}_{1}\cdot\mathbf{s}_{2}=\cos\theta_{12}=-1/3$,
that is, for $\theta_{12}\approx110\lyxmathsym{\textdegree}$, the
total spin is frozen.

\begin{figure}
\begin{centering}
\includegraphics[width=9cm]{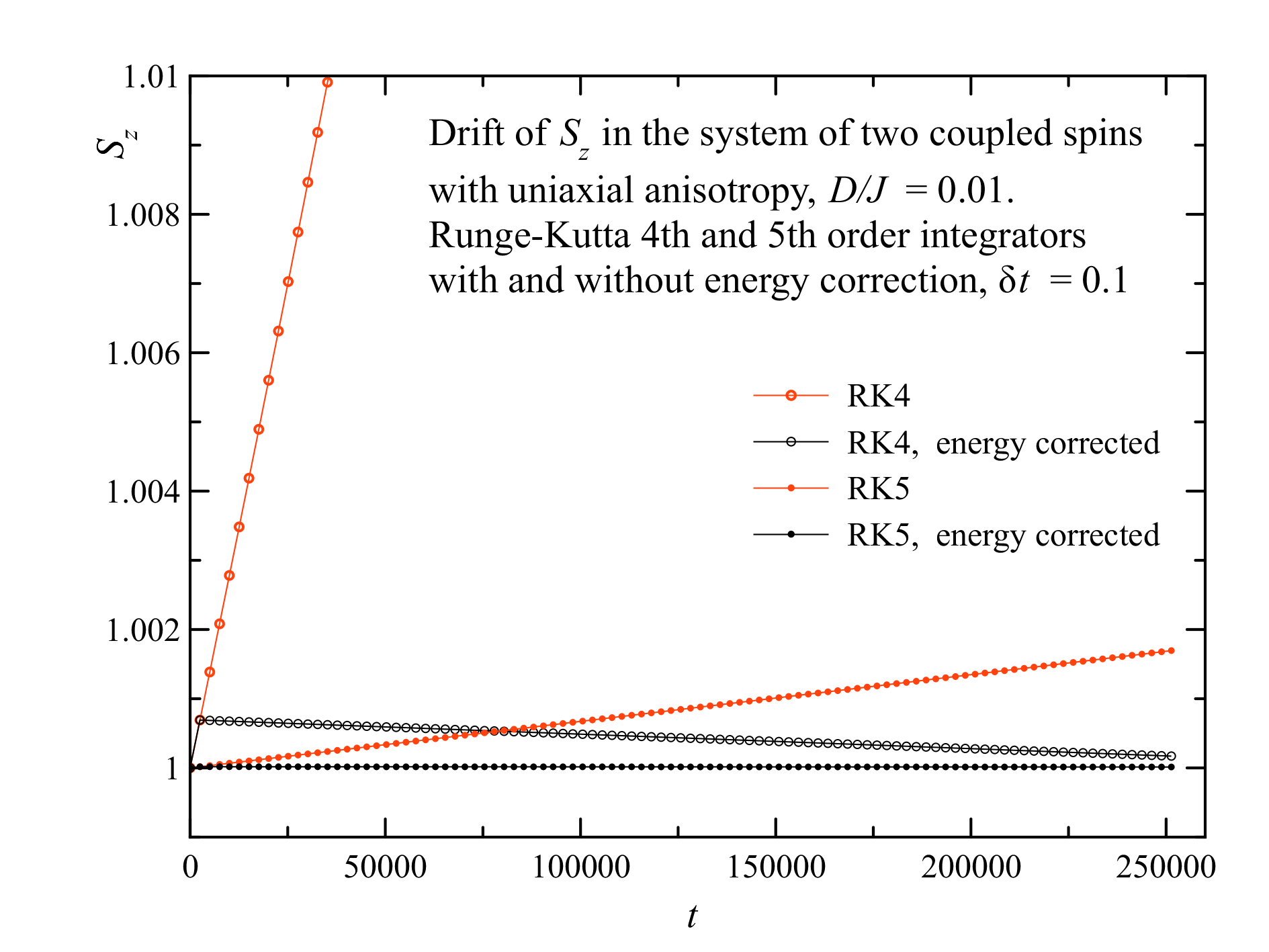}
\par\end{centering}
\caption{Drift of $S_{z}$ in the system of two coupled spins with uniaxial
anisotropy $D/J=0.01$, computed with Runge-Kutta fourth- and fifth-order
integrators with and without energy correction, with $\delta t=0.1$
\label{Fig-Sz_vs_t_D=00003D0p01_hStep=00003D0p1_zoom}}

\end{figure}

For the initial spin state $\left\{ \mathbf{s}_{1},\mathbf{s}_{2}\right\} =\left\{ \left(0,0,1\right),\left(1,0,0\right)\right\} $
one has $\mathbf{s}_{1}\cdot\mathbf{s}_{2}=0$, and $S_{z}=1$, and
Eq. (\ref{Omega_res}) yields
\begin{equation}
\hbar\Omega=\frac{D}{4}.\label{Omega_case_1}
\end{equation}

For the initial state $\left\{ \left(-1/\sqrt{2},0,1/\sqrt{2}\right),\left(1,0,0\right)\right\} $
one has $\mathbf{s}_{1}\cdot\mathbf{s}_{2}=-1/\sqrt{2}$, and $S_{z}=1/\sqrt{2}$,
and Eq. (\ref{Omega_res}) yields
\begin{equation}
\hbar\Omega=\frac{D}{4\sqrt{2}}\frac{\sqrt{2}-3}{\sqrt{2}-1}\simeq-0.677D.\label{Omega_case_2}
\end{equation}

Figure \ref{Fig-Sxyz_vs_t_D=00003D0p01_hStep=00003D0p1_both_cases}
shows the numerical solution of the system of equations (\ref{s1s2_Eqs})
in both cases above for $D/J=0.01$. For such a small anisotropy,
the curves for $S_{x}$ and $S_{y}$ are visibly perfect sinusoidals,
while $S_{z}$ is a straight line. In the first case, the period is
$T=2513$ (in units of $\hbar/J$), in perfect accordance with the
value $T=2\pi/\Omega=2513$ following from Eq. (\ref{Omega_case_1}).
In the second case, the precession goes in the other direction with
the period $T=936$, in reasonable accordance with the result $T=928$
of Eq. (\ref{Omega_case_2}) (in this case, the approximation made
in the derivation of $\Omega$ works less well). The fast motion of
the difference spin $\boldsymbol{\sigma}$ is not seen in this figure.
For larger anisotropies, such as $D/J\gtrsim0.1$, the numerical solution
shows a more complicated behavior with both types of motion.

This toy model is well suited for checking the methods of integrating
equations of motion for classical spin systems. As in the real systems,
here there is fast precession of spins around each other with the
exchange frequency $\omega_{\mathrm{ex}}\sim J/\hbar$ that in real
systems becomes important at high excitation, in particular, at elevated
temperatures. At the same time, there is a slow motion of the observed
macroscipic quantities, driven by the interactions much weaker than
the exchange. Although the latter are of interest, the integration
step $\delta t$ in the numerical solution is dictated by the fast
motion and is typically $\delta t\sim0.1$ in the units of $\hbar/J$.
This leads to very long computations even for physically fast processes.
In such computations, numerical errors tend to accumulate. This is
why the energy-conserving symplectic integrators have become widely
accepted.

To demonstrate that the metod of energy correcting proposed above
is efficient in long computations using RK4 and RK5 ODE solvers, computations
on the toy model with $D/J=0.01$ and the initial spin configuration
$\left\{ \mathbf{s}_{1},\mathbf{s}_{2}\right\} =\left\{ \left(0,0,1\right),\left(1,0,0\right)\right\} $
were performed over 100 periods of the precession of the total spin,
$T=2\pi/\Omega$ specified by Eq. (\ref{Omega_case_1}).

\begin{figure}
\begin{centering}
\includegraphics[width=9cm]{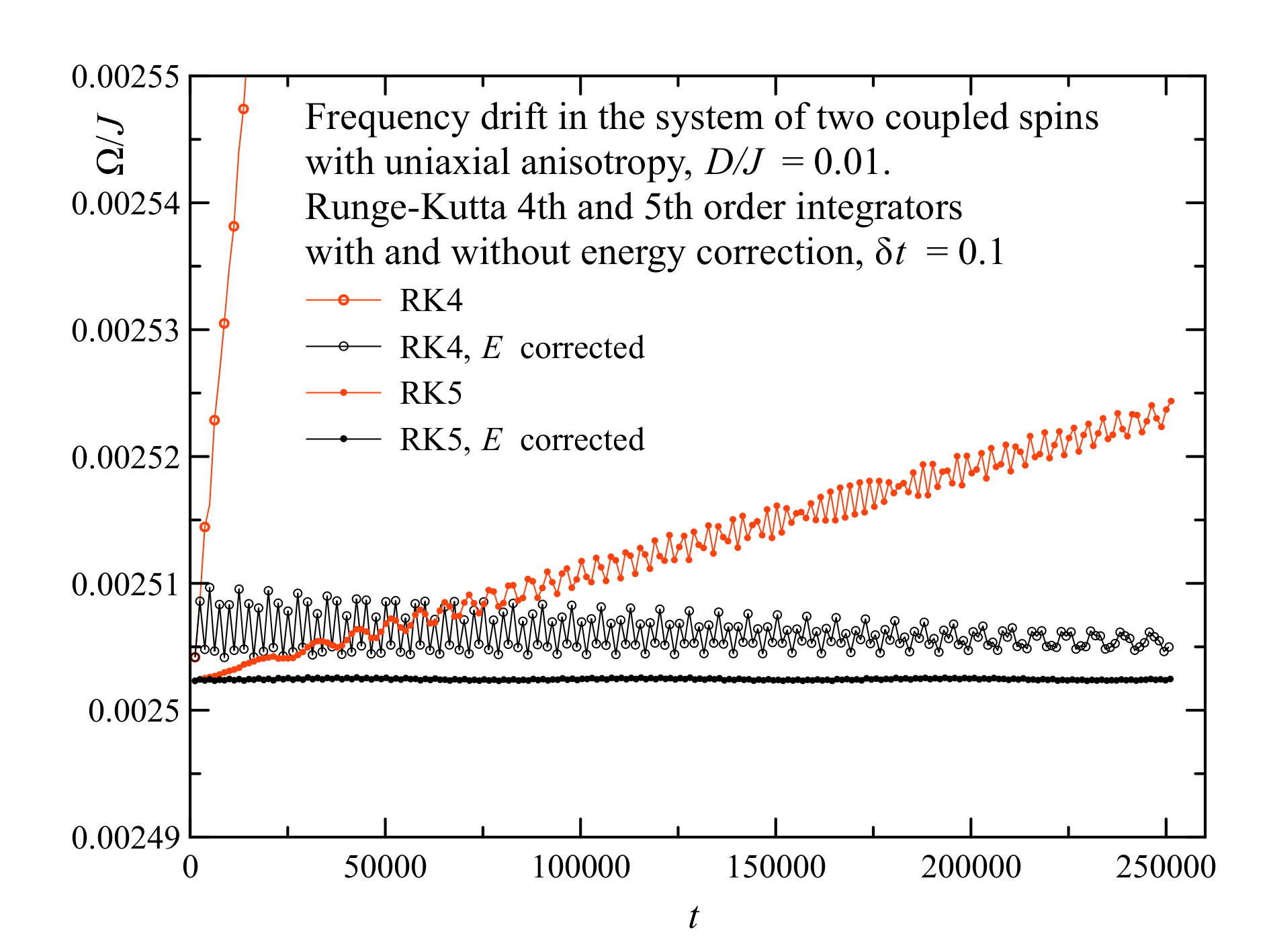}
\par\end{centering}
\caption{Frequency drift in the system of two coupled spins with uniaxial anisotropy
$D/J=0.01$, computed with Runge-Kutta fourth- and fifth-order integrators
with and without energy correction, with $\delta t=0.1$ \label{Fig-Frequency_vs_t_D=00003D0p01_hStep=00003D0p1_zoom}}

\end{figure}

\begin{figure}
\begin{centering}
\includegraphics[width=9cm]{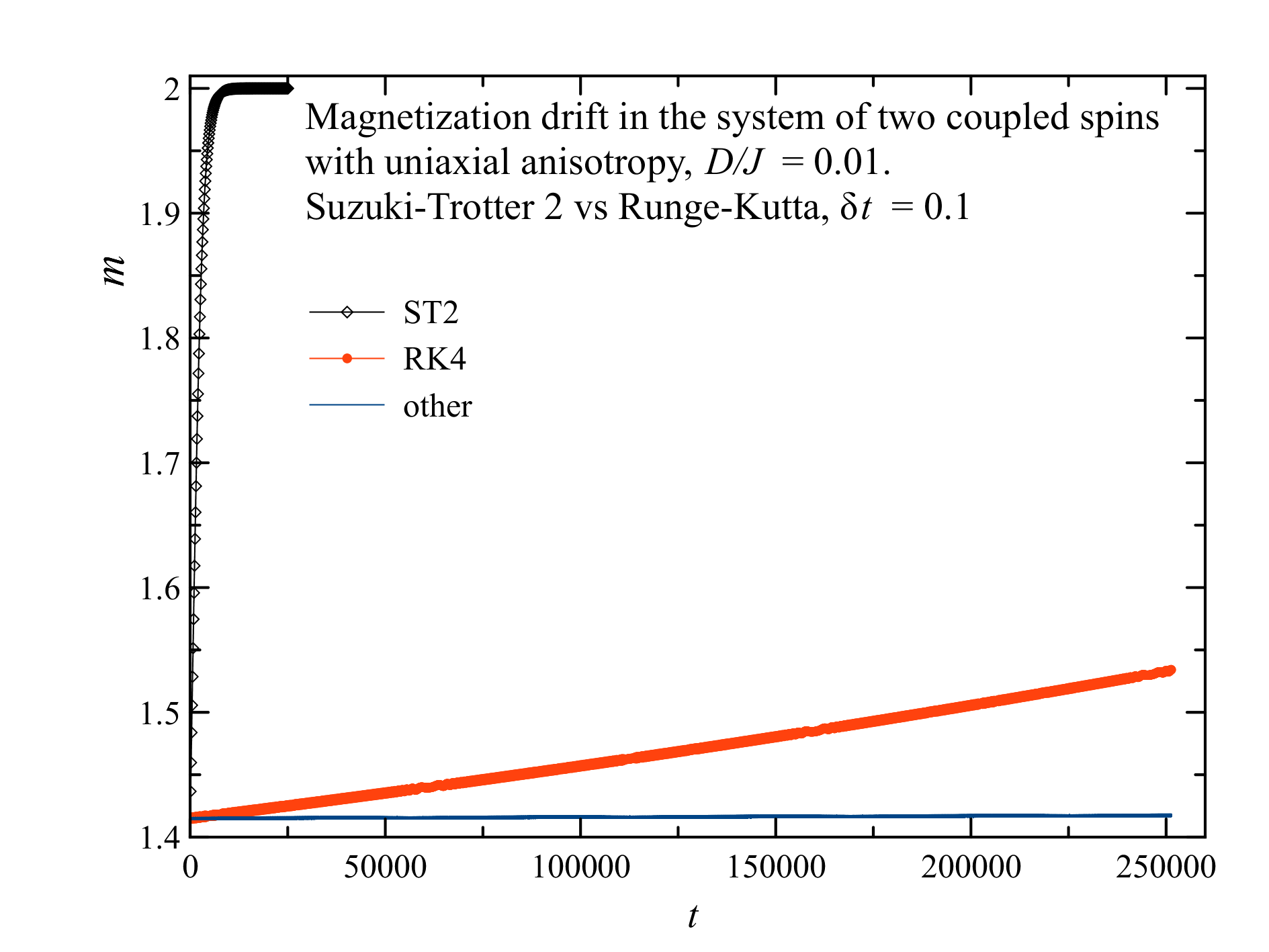}\caption{Magnetization drift in the system of two coupled spins with uniaxial
anisotropy $D/J=0.01$, computed with the ST2 and Runge-Kutta methods,
with $\delta t=0.1$\label{Fig-m_vs_t_D=00003D0p01_ST_vs_RK}}
\par\end{centering}
\end{figure}

Figure \ref{Fig-DeltaE_vs_t_D=00003D0p01_hStep=00003D0p1_ST2_vs_RK}
shows the energy drift computed with the corrected and uncorrected
RK4 and RK5 methods, as well as with the second-order Suzuki-Trotter
decomposition for comparison (all spins are rotated sequentially by
half-angles around their effective fields and then the same in the
opposite order \citep{basvergarkac17jpcm}, with no attempt to solve
the problem of a non-constant effective field by iterations \citep{krebunlan98,lanbunevekretsa00}).
In the case of ST2, the energy decreases very fast and saturates at
$\Delta E/J=-1$, which corresponds to the angle between the spins
decreasing from its initial value $90\lyxmathsym{\textdegree}$ to
zero (see Fig. \ref{Fig-m_vs_t_D=00003D0p01_ST_vs_RK}). This confirms
an extreme inaccuracy of the ST2 method for systems with uniaxial
anisotropy that in this case acts as an effective damping. As said
in the Introduction, ST2 straightforwardly applied to such systems
has, in fact, a step error $\delta t^{2}$, which is inferior to that
of RK4 having a step error $\delta t^{5}$. Still, over this huge
integration time, the RK4 energy drift is also significant, $\varDelta E/J\simeq-0.0874$.

Figure \ref{Fig-DeltaE_vs_t_periods_D=00003D0p01_hStep=00003D0p1_zoom}
shows a close-up of the energy drift. Correcting the energy every
half-period $T/2$ of the slow precession with the RK4 integrator
yields a constant energy deviation $\varDelta E/J\simeq-0.7\times10^{-3}$
that is not that bad, especially as it is not growing with time. This
energy deviation accumulates over the time $T/2$, after which the
energy each time returns to its target value. As here $T/2\simeq1257$
and the integration step is $0.1$, energy corrections are performed
extremely rarely and in fact can be done much more frequently, further
reducing the energy deviation. Uncorrected RK5 computation has much
better accuracy than the uncorrected RK4 one, as can be seen in Fig.
\ref{Fig-DeltaE_vs_t_periods_D=00003D0p01_hStep=00003D0p1_zoom}.
Energy corrections for RK5 make errors in the energy invisible on
this scale.

Energy plays a profound role in the dynamics, affecting other physical
quantities, as the value of the energy defines the region of the phase
space that the system is allowed to visit. The negative energy drift
in the present uncorrected computations causes the spins to move closer
to the anisotropy axis. As a result, there is a positive drift in
the integral of motion $S_{z}$ and a positive drift in the slow precession
frequency. Figure \ref{Fig-Sz_vs_t_D=00003D0p01_hStep=00003D0p1_zoom}
shows the numerical results for $S_{z}$. Indeed, for the uncorrected
RK4 and RK5 solvers $S_{z}$ increases, and for RK4 this increase
is very pronounced ($S_{z}=1.0834$ at the end of the integration
interval). RK4 with energy correction yields a small $S_{z}$ drift,
even decreasing with time. Using RK5 with energy correction makes
$S_{z}$ errors invisible on this scale.

The frequency drift shown in Fig. \ref{Fig-Frequency_vs_t_D=00003D0p01_hStep=00003D0p1_zoom}
is similar to the $S_{z}$ drift. In the uncorrected RK4 computation,
the frequency $\Omega$ becomes $0.0035J$ at the end of the integration
time, which is a huge deviation from the correct value of $\Omega$.
The accurate numerical calculation yields $\Omega$ slightly higher
than the theoretical value $0.0025J$. The reason for this is that
Eq. (\ref{Omega_res}) is an approximate analytical result and there
should be corrections to it.

Finally, the time dependence of the length of the total spin $m=\left|\mathbf{S}\right|=\sqrt{2\left(1+\mathbf{s}_{1}\cdot\mathbf{s}_{2}\right)}$
is shown in Fig. \ref{Fig-m_vs_t_D=00003D0p01_ST_vs_RK}. Applying
ST2 makes the two spins, initially perpendicular to each other, become
parallel, reducing the energy (see Fig. \ref{Fig-DeltaE_vs_t_D=00003D0p01_hStep=00003D0p1_ST2_vs_RK}).
The drift of $m$ in the uncorrected RK4 computation is also substantial.
In the uncorrected RK5 computation, there is also a positive drift
of $m$; however, invisible on this scale. In the corrected RK4 and
RK5 computations, there is no $m$ drift but, upon zooming in, one
can see fast oscillations with a very small amplitude, as $m$ is
not conserved in this model.

\section{Energy correction in systems with damping and pumping}

If the spin system is damped, $\alpha>0$, and $T=0$, so that there
is no thermal agitation, the system will relax fast to its ground
state, so that no significant error will accumulate during the relaxation,
if the integrator is accurate enough. The problem of accumulation
of errors arizes in the case of continuous pumping, which causes nontrivial
dynamics during an extended time. The absorbed and dissipated energies
\begin{eqnarray}
E_{\mathrm{abs}}(t) & = & \intop_{0}^{t}dt'P_{\mathrm{abs}}(t'),\nonumber \\
E_{\mathrm{diss}}(t) & = & \intop_{0}^{t}dt'P_{\mathrm{diss}}(t'),\label{E_abs_diss}
\end{eqnarray}
where $P_{\mathrm{abs}}$ and $P_{\mathrm{diss}}$ are given by Eq.
(\ref{P_abs_diss}), are robust quantities. The contributions to them
obtained by the integration up to the current time are recorded and
do not change any more. The only change of $E_{\mathrm{abs}}(t)$
and $E_{\mathrm{diss}}(t)$ is due to the further evolution. On the
contrary, the energy of the system $\mathcal{H}$ is subject to drift
because of the accumulation of errors in the numerical solution of
the system's dynamics that is on during the whole computation. If
$\mathcal{H}$ strongly deviates from its accurate value, the state
of the system changes so that $P_{\mathrm{abs}}$ and $P_{\mathrm{diss}}$
become wrong, and the whole computation breaks down. The key to the
correct computation is in correcting the energy change $\Delta E(t)\equiv\mathcal{H}(t)-\mathcal{H}(0)$
so that it satisfies the energy balance condition (\ref{E_balance})
integrated over the time, i.e.,
\begin{equation}
\Delta E(t)=E_{\mathrm{abs}}(t)-E_{\mathrm{diss}}(t).
\end{equation}
Thus, in Eq. (\ref{deltaE}) $E_{\mathrm{target}}=\mathcal{H}(0)+E_{\mathrm{abs}}(t)-E_{\mathrm{diss}}(t)$,
and the required energy correction becomes
\begin{equation}
\delta E=E_{\mathrm{abs}}(t)-E_{\mathrm{diss}}(t)-\Delta E(t).
\end{equation}

Such a strategy was applied in recent work \citep{garchu21_absorption_eprint},
where the absorption of microwave energy in a large system of classical
spins with random anisotropy was studied. Although no phenomenological
damping was included, as the system of many interacting spins has
its own internal damping, this work illustrates well the power of
the energy-correcting procedure. With accurate numerical integration
in Eqs. (\ref{E_abs_diss}) and the energys correction, the evolution
of the system can be traced up to unlimited times. With the use of
the RK 5 integrator, the integration was performed with a time step
of $\delta t=0.1$ up to $t=100000$ in units of $\hbar/J$.

Of course, there will be some error accumulation because of inaccuracies
of the numerical integration in the formulas for $E_{\mathrm{abs}}(t)$
and $E_{\mathrm{diss}}(t)$ {[}Eqs. (\ref{E_abs_diss}){]}. However,
these errors just slightly renormalize the pumping and damping and
cannot result in any drastic effects.

If the pumping and damping are strong, the non-conservation of the
system's energy resulting from the inaccuracy of the ODE solver is
not very important as it only slightly shifts the tight and fast establishing
balance between pumping and damping. The energy correction becomes
necessary when pumping and damping are weak, so the process is so
long that the accumulated errors in the system\textquoteright s energy
due to the inaccuracy of the ODE solver become noticeable.

In the case of nonzero temperature $T$, one should keep in mind that
the energy-balance equation (\ref{E_balance}) is averaged over the
fluctuations of the phenomenological stochastic fields. The equations
to solve are stochastic equations, the solution of which is fluctuating
around the average value at a given time. Also, the system's energy
is fluctuating and because of this it cannot be corrected as was described
above. On the other hand, thermal agitation tends to restore the average
system's energy in a natural way. The result of numerical errors will
be just a small deviation of the dynamical spin temperature $T_{S}$
from the temperature of the bath $T$ and there will be no error accumulation.
Thus, having a sufficiently accurate ODE solver, one can forget about
the energy correction for $T>0$.

An efficient method of solving the stochastic Landau-Lifshitz-Langevin
equation for classical spins, especially in the realistic case of
weak damping $\alpha\ll1$, is replacing the continuous noise by a
pulse noise \citep{gar17pre} acting at time intervals $\Delta t$.
Within these intervals, the motion of the system is noiseless and
it can be solved by high-accuracy ODE solvers such as RK4 or RK5.
This is an importand advantage in comparison with the standard approach
using the original continuous noise that requires using the low-accuracy
Heun (a variant of RK2) integrator with a very small integration step.
Within the pulse-noise scheme, one can implement the energy correction
at the end of each interval $\Delta t$ to ensure a proper energy
behavior within this interval.

\section{Discussion}

\label{sec:Discussion}

It was shown that mainstream ODE solvers, not explicitly conserving
the energy for conservative classical-spin systems, can be used for
solving the equations of motions for spins over very long times, if
the energy-correction procedure is employed in the algorithm. This
procedure, executed from time to time, returns the value of the energy
of the spin system to its target value computed from the initial energy
and the energy injected into and dissipated in the system, which are
not subject to drift. In particular, one can use the classical fourth-order
Runge-Kutta solver or the Butcher's fifth-order Runge-Kutta solver.
For many-spin systems, these solvers can be written in the vector
form so that the code looks like that for one differential equation.
Correcting the energy also makes other computed physical quantities
more accurate.

The energy-correction method can be implemented both for the pure
spin dynamics with the phenomenological damping and Langevin stochastic
fields simulating the heat bath (if the pulse-noise model \citep{gar17pre}
is used) and for the combined spin-lattice dynamics. In both cases,
the target energy of the spin system can be computed.

The method is especially useful for spin systems with single-site
anisotropy for which the popular symplectic integrators based on the
Suzuki-Trotter decomposition of exponential operators do not conserve
energy and thus become inefficient. Even in the absence of single-site
interactions, mainstream methods with energy correction are competitive
with symplectic methods. For instance, the second-order Runge-Kutta
(RK2) solver makes two function evaluations per integration step,
while the most used second-order Suzuki-Trotter solver, ST2, also
makes two effective function evaluations per step, only it does it
sequentially for all spins. The RK4 solver has the fourth order of
accuracy and makes four evaluations per step but the ST4 solver makes
$5\times2=10$ \citep{krebunlan98,lanbunevekretsa00} effective function
evaluations per step. It is inferior to Butcher's RK5 that makes six
function evaluations per step.

How frequently energy corrections have to be done depends on the error
accumulated during the time between the corrections. The latter depends
on the particular problem and on the integration step. Thus, before
the definitive computation is run, different variants have to be tested.

Considering the energy balance in classical spin systems allowed us
to obtain the formula for the dynamic spin temperature in the presence
of single-site anisotropy, generalizing the previously obtained results
for different-site interactions. This formula is useful in studying
spin dynamics.

\section*{Acknowledgements}

This work was supported by Grant No. FA9550-20-1-0299 funded by the
Air Force Office of Scientific Research. The author thanks E. M. Chudnovsky
for discussing this research at every stage.

\section*{Appendix}
\begin{flushleft}
In the term $\mathbf{H}_{\mathrm{eff},i}\cdot\left(\mathbf{s}_{i}\times\boldsymbol{\zeta}_{i}\right)$
in Eq. (\ref{E_balance_0}), the Langevin field $\boldsymbol{\zeta}_{i}$
directly correlates with $\mathbf{s}_{i}$ and, in the presence of
single-site interactions, with $\mathbf{H}_{\mathrm{eff},i}$. Thus,
averaging over realizations of $\boldsymbol{\zeta}_{i}$, one has
to calculate two terms:
\begin{equation}
\left\langle \mathbf{H}_{\mathrm{eff},i}\cdot\left(\mathbf{s}_{i}\times\boldsymbol{\zeta}_{i}\right)\right\rangle =A+B,
\end{equation}
where
\begin{equation}
A\equiv\mathbf{H}_{\mathrm{eff},i}\cdot\left\langle \mathbf{s}_{i}\times\boldsymbol{\zeta}_{i}\right\rangle ,\qquad B\equiv\left\langle \boldsymbol{\zeta}_{i}\cdot(\mathbf{H}_{\mathrm{eff},i}\right\rangle \times\mathbf{s}_{i}).
\end{equation}
One can use the implicit solution
\begin{equation}
s_{i\alpha}(t)=\frac{1}{\hbar}\intop_{t_{0}}^{t}dt'e_{\alpha\beta\gamma}s_{i\beta}(t')\zeta_{i\gamma}(t')+\ldots\label{s_integral}
\end{equation}
for the dependence of $\mathbf{s}_{i}$ on $\boldsymbol{\zeta}_{i}$
that follows from Eq. (\ref{LLL}). Then in $A$, one has
\begin{eqnarray}
\left(\mathbf{s}_{i}(t)\times\boldsymbol{\zeta}_{i}(t)\right)_{\alpha}=e_{\alpha\mu\nu}s_{i\mu}(t)\zeta_{i\nu}(t)\nonumber \\
=\frac{1}{\hbar}\intop_{t_{0}}^{t}dt'e_{\alpha\mu\nu}\zeta_{i\nu}(t)e_{\mu\beta\gamma}s_{i\beta}(t')\zeta_{i\gamma}(t')+\ldots
\end{eqnarray}
Using the identity $e_{\mu\nu\alpha}e_{\mu\beta\gamma}=\delta_{\nu\beta}\delta_{\alpha\gamma}-\delta_{\nu\gamma}\delta_{\alpha\beta}$,
one can rewrite this as
\begin{eqnarray}
\frac{1}{\hbar}\intop_{t_{0}}^{t}dt'\left(\delta_{\nu\beta}\delta_{\alpha\gamma}-\delta_{\nu\gamma}\delta_{\alpha\beta}\right)\zeta_{i\nu}(t)s_{i\beta}(t')\zeta_{i\gamma}(t') & =\nonumber \\
\frac{1}{\hbar}\intop_{t_{0}}^{t}dt'\left[\zeta_{i\beta}(t)s_{i\beta}(t')\zeta_{i\alpha}(t')-\zeta_{i\gamma}(t)s_{i\alpha}(t')\zeta_{i\gamma}(t')\right].
\end{eqnarray}
Here the correlator of the Langevin fields is equal to 1/2 of the
value given by Eq. (\ref{Langevin-correlator}) as $t'=t$ is the
upper limit of the integral. Thus one obtains
\begin{equation}
\left\langle \left(\mathbf{s}_{i}(t)\times\boldsymbol{\zeta}_{i}(t)\right)_{\alpha}\right\rangle =\alpha Ts_{i\alpha}(t)-3\alpha Ts_{i\alpha}(t)=-2\alpha Ts_{i\alpha}(t)
\end{equation}
and
\begin{equation}
A\equiv\mathbf{H}_{\mathrm{eff},i}\cdot\left\langle \left(\mathbf{s}_{i}\times\boldsymbol{\zeta}_{i}\right)\right\rangle =-2\alpha T\left(\mathbf{H}_{\mathrm{eff},i}\cdot\mathbf{s}_{i}\right).
\end{equation}
\par\end{flushleft}

Let us calculate now the $B$-term. The contribution to $B$ comes
from the uniaxial anisotropy, see Eq. (\ref{Heff}):
\begin{equation}
B\equiv\left\langle \boldsymbol{\zeta}_{i}\cdot(\mathbf{H}_{\mathrm{eff},i}\right\rangle \times\mathbf{s}_{i})=D\left\langle \left(\mathbf{n}_{i}\cdot\mathbf{s}_{i}\right)\boldsymbol{\zeta}_{i}\right\rangle \cdot\left(\mathbf{n}_{i}\times\mathbf{s}_{i}\right).
\end{equation}
Similarly to the above, one writes
\begin{eqnarray}
\left\langle \left(\mathbf{n}_{i}\cdot\mathbf{s}_{i}\right)\zeta_{i\alpha}\right\rangle  & = & \frac{1}{\hbar}\left\langle \intop_{t_{0}}^{t}dt'n_{i\nu}e_{\nu\eta\gamma}s_{i\eta}(t')\zeta_{i\gamma}(t')\zeta_{i\alpha}(t)\right\rangle \nonumber \\
 & = & \alpha Tn_{i\nu}e_{\nu\eta\alpha}s_{i\eta}.
\end{eqnarray}
That is,
\begin{equation}
\left\langle \left(\mathbf{n}_{i}\cdot\mathbf{s}_{i}\right)\boldsymbol{\zeta}_{i}\right\rangle =\alpha T\left(\mathbf{n}_{i}\times\mathbf{s}_{i}\right)
\end{equation}
and
\begin{equation}
B=\alpha TD\left(\mathbf{n}_{i}\times\mathbf{s}_{i}\right)^{2}=\alpha TD\left[1-\left(\mathbf{n}_{i}\cdot\mathbf{s}_{i}\right)^{2}\right].
\end{equation}
Finally, adding $A$ and $B$ and grouping the terms containing the
uniaxial anisotropy, one obtains
\begin{eqnarray}
\left\langle \mathbf{H}_{\mathrm{eff},i}\cdot\left(\mathbf{s}_{i}\times\boldsymbol{\zeta}_{i}\right)\right\rangle  & = & -2\alpha T\left(\widetilde{\mathbf{H}}_{\mathrm{eff},i}\cdot\mathbf{s}_{i}\right)\nonumber \\
 &  & -\alpha TD\left[3\left(\mathbf{n}_{i}\cdot\mathbf{s}_{i}\right)^{2}-1\right],
\end{eqnarray}
where $\widetilde{\mathbf{H}}_{\mathrm{eff},i}=\sum_{j}J_{ij}\mathbf{s}_{j}$
is the effective field without the anisotropy term.


\newcommand{\noopsort}[1]{} \newcommand{\printfirst}[2]{#1}
  \newcommand{\singleletter}[1]{#1} \newcommand{\switchargs}[2]{#2#1}

\end{document}